# The Impacts of Low/No-Code Development on Digital Transformation and Software Development

Zhaohang Yan
Department of Computer & Mathematical Sciences
University of Toronto
Toronto, Canada
zhaohang.yan@mail.utoronto.ca

*Abstract*—**Low/No-code development is a software development method that provides users with a platform for visually creating applications with little or no coding. Companies and organizations need software applications and information systems for various business purposes like management in the technology era. Low/No-code development gives non-IT professionals a convenient tool for rapidly building simple business applications they need without or with little coding. In this paper, we explored the benefits & limitations of Low/No-Code development and modern Low/No-Code development platforms in the industry. In addition, we analyzed how it can be improved and prospected the impacts of Low/No-Code development on society and related industries in the future. In conclusion, we find that Low/No-code development is a promising trend that can significantly impact future software development and digital transformation.**

## I. INTRODUCTION

With the development of information technology and digitalization trends, companies and organizations require robust tools to respond to dynamic & complex market environments and requirements (Sanchis et al., 2020) [1]. In his journal Software development trends 2021, McLean (2021) defines Low/No-Code development as a sort of visual software development that developers can use to drag & drop and connect components for building mobile or web applications [4]. Such a component-based approach empowers professional developers to deftly create applications without any code (McLean, 2021) [4]. In 2014, Richardson and Rymer (2014) first brought in the phrase "low code" to the public in their report titled "New Development Platforms Emerge For Customer-Facing Applications," published in Forrester Research, in which they claim that comparing to demanding hand-coding, enterprises and organizations prefer low-code substitutes for swift, continual, and test-and-learn development and delivery [2].

The result of an investigation conducted by Richardson and Rymer (2016) illustrated that low code development platforms could help enterprises and organizations expedite the development and delivery of applications by 5 to 10 times [3]. Gartner, a technology research and consulting company, forecasts that Low-Code Development will be used for 65% of application development projects by 2024 (Duffy, 2019) [5]. Therefore, it is crucial for the public to understand LNCD comprehensively when LNCD is rapidly adopted and embraced by enterprises and organizations.

Nevertheless, though Low/No-Code Development has many good for application development, some enterprises do not adopt it over conventional software development process due to the dearth of knowledge about Low/No-Code Development Platforms (LNCDP), fear of vendor lock-in (i.e., fear of being dependent on an LNCDP vendor), the scarcity of confidence that LNCDP can develop desired applications, worries about the security, scalability, and inflexibility issues (OutSystems, 2019) [7].

However, some of the concerns about LNCDP, such as inflexibility, scalability, and security, were defended or proved to be manageable by Warren from OutSystems (2018) [8]. Alsaadi et al. (2021) analyzed leading LNCDP developers such as OutSystems, Microsoft, Salesforce, Mendix, and Appian [11]. We will analyze the features and functionalities of these platforms and focus on the insight we get by looking at trends and patterns in the Low/No-Code industry.

In the last section of this paper, we prospect how Low/No-Code development will affect future software development and digital transformation by analyzing opposite opinions.

We conclude that even if Low/No-Code Development emerged only a few years ago, the rapid growth of its adoption in the industry indicates an optimistic prospect. Based on our analysis in each section, Low/No-Code Development will cause a turnaround in the software development industry and conduce digital transformation.

## II. BENEFITS AND LIMITATIONS OF LOW/NO-CODE DEVELOPMENT

In today's society, it becomes more critical for enterprises and organizations to adapt to fast-changing both external and internal environments (OutSystems, 2019) [7]. As defined by

Alsaadi et al. (2021), digital transformation usually refers to digitalizing paper-based documents and storing them in a digital form [11]. It transforms the business processes from time-consuming and labour-intensive manual processes to automated agile digital processes (Alsaadi et al., 2021) [11]. According to Outsystems (2019), digital transformation has been playing a significant role in business strategy for a few years [7]. Organizations leverage digital transformation to grasp new opportunities, to satisfy customers' complex and various needs, and to circumvent being defeated by competitors that become more adaptive and flexible (OutSystems, 2019) [7]. Outsystems (2019) investigated over 3,300 IT professionals from 6 different continents, and Figure 1 shows the results regarding business digital transformation progress. According to Outsystems' (2019) criteria, Level 1 Unaware is doing almost nothing for digital transformation; Level 2 Isolated is the preliminary stage of digital transformation, and Level 6 is the most robust stage [7]. We can conclude from the chart that the majority (95%) of the respondents' organizations were more or less making efforts for digital transformation.

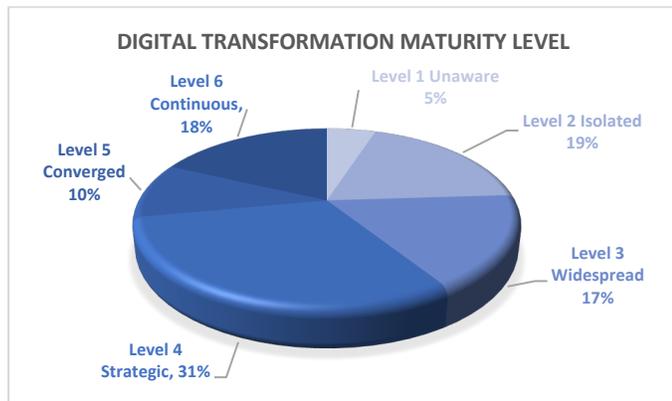

(Figure 1. Digital Transformation Maturity Level. Data source: Outsystems (2019) [7])

Outsystems (2019) also surveyed Applications Planned for Development in 2019 and Application Development Time. Figure 2 shows that 23% of the respondents planned to develop and deliver over 50 software applications. Figure 3 shows that majority of the applications are required over three months. We conclude from these two graphs that during digital transformation, there will be numerous and continuous demands for new application development and delivery, which can be time-consuming. Therefore, organizations that plan to or are digital transforming need a tool to improve development & delivery efficiency. Low/No-code development is one of the rising approaches to boost delivery speed. Figure 4 illustrates the percentage of Low/No-code adoption in the IT strategy of respondent organizations, where we can see that over half of them will start or are utilizing Low/No-code development. The study of Richardson and Rymer (2016) indicates that Low/No-code development can speed up the applications development & delivery by 5 to 10 times [3]. According to Duffy (2019), Gartner predicted that, by 2024, Low-Code Development would take on 65% of application development tasks [5]. For the Digital Transformation Maturity (DTM) we mentioned in Figure 1, the average DTM score of Low-Code practitioners was 16% higher than that of industry peers who did not use Low-Code (Outsystems, 2019) [7]. Moreover, 37% of businesses that practiced Low-Code development were satisfied with the software application delivery rate, while only 26% of businesses that did not practice Low-Code were content with their release rate (Outsystems, 2019) [7]. Therefore, it is important to learn more comprehensively about Low/No-code development before using it during digital transformation. In this section, we explore both benefits and limitations of Low/No-code development for developers and organizations to pounder if it is felicitous with their digital transformation strategy.

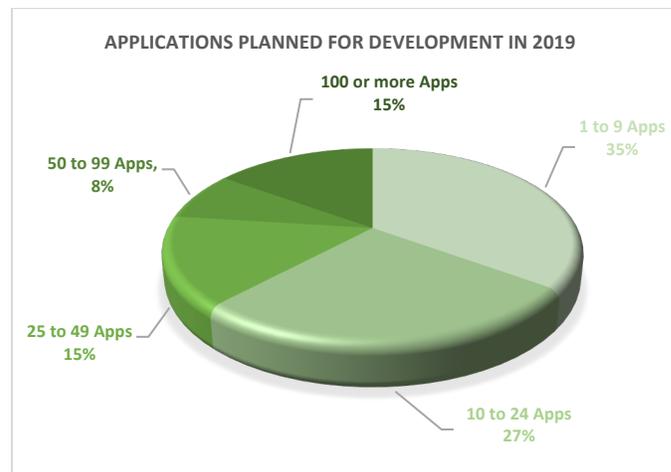

(Figure 2. Applications Planned for Development. Data source: Outsystems (2019) [7])

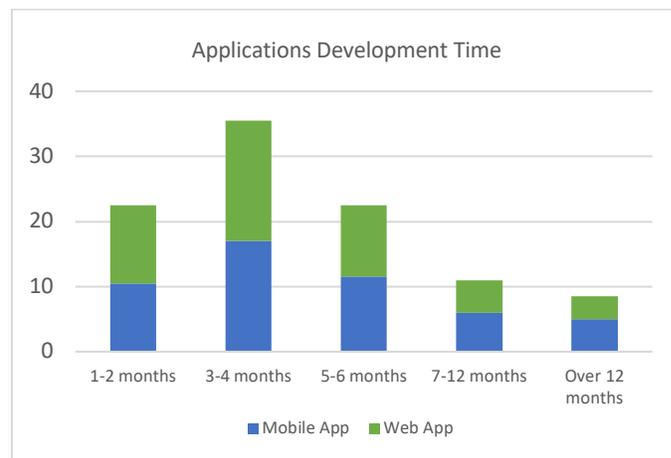

(Figure 3. Applications Development Time. Data source: Outsystems (2019) [7])

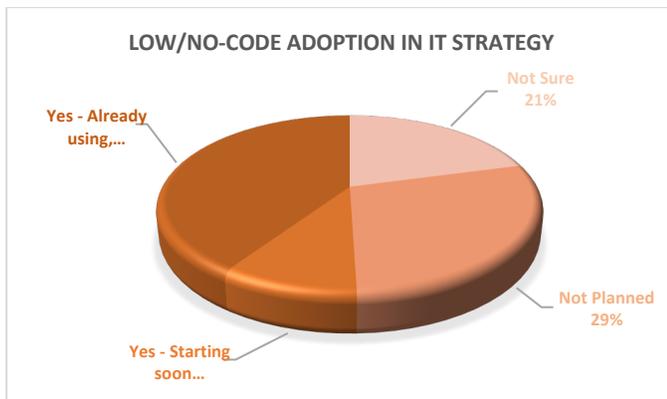

(Figure 4. Low/No-code Adoption in IT Strategy. Data source: Outsystems (2019) [7])

*A.* BENEFITS

There are many goals that developers want to achieve using Low-code development. Outsystems (2019) collected top reasons for utilizing Low-code development, and the first three were Expedite digital innovation/transformation (66%), Improve responsiveness to the business (66%), Decrease dependency on the technical skill that is hard to hire (45%) [7]. Besides these general responses, we expand several main benefits that Low/No-code development can produce in this sub-section.

a. Swiftness

By the nature of Low/No-code development, Low/No-code development platforms allow users to build new applications visually, which accelerate development by providing businesses with immediate prototyping new customer requirements and testing of the functionalities (Richardson & Rymer, 2016) [3]. Richardson and Rymer (2016) also argued that coding a new software application is still a demanding and time-consuming task, even though practicing Agile or Lean software development process [3]. Modifying a customized application with numerous lines of code is also complicated and inefficient (Richardson & Rymer, 2016) [3]. Most of the enterprises using Low-code platforms reported that Low-code development conduced to their release of applications and speeded it up by 5 to 10 times (Richardson & Rymer, 2016) [3]. As presented in Outsystems (2019) survey, Low-Code practitioners could release 68% of their web applications in 4 months, while non-Low-Code users could only deliver 57% of their web applications using the same amount of time [7]. For mobile apps, it presented a similar pattern where Low-Code practitioners could deliver 64% of the apps in 4 months, but those who did not use Low-Code could only deliver 49% of apps in 4 months (OutSystems, 2019) [7].

b. Citizen Development

Citizen developers are non-professional developers authorized by the IT department to use approved tools for application development & delivery (OutSystems, 2019) [7]. Though citizen developers lack programming skills, they have abundant experience and knowledge in other fields; since programming skills are hardly needed when using Low-Code, citizen developers could use the knowledge and experience to develop applications that meet customers' needs better (Alsaadi et al., 2021) [11]. When the enterprises have a dearth of IT talents but with a myriad of demands of new web/mobile applications from various departments, citizen developers in each department could use Low/No-Code development platforms to build needed business applications for their departments (Richardson & Rymer, 2016) [3]. It not only satisfies departments' needs but also releases IT professionals from tedious low-level application development details. IT professionals could then focus on more complex and essential business application requirements. Outsystems (2019) assessed the administration of citizen developers and found that 37% of Low-Code practitioners could administrate and manage citizen development attentively, which was 9% higher than that of non-Low-Code users [7]. Therefore, Low/No-Code development improves citizen development on the developer and administrator sides.

c. Security

During digital transformation, the limited number of IT talents makes it hard for IT professionals to manage a mass of demands of business applications (Stangarone, J., 2019) [15]. In such a situation, people who cannot get solutions from the IT department or IT professionals would look for third-party solutions without the oversight of the IT department, which is known as "Shadow IT" (Stangarone, J., 2019) [15]. Shadow IT can endanger an organization's IT security and privacy since the organization will have no knowledge and control of the applications such as implementation details, data management, and IT security threats (OutSystems, 2019) [7]. Low/No-Code development platforms, authorized by IT personnel or IT department, can help solve the risks brought by Shadow IT. Users can build needed solutions on the platforms without disturbing IT personnel constantly. Since the IT department controls the data and applications, security and privacy safety are guaranteed (Stangarone, J., 2019) [15]. Security concerns about citizen development were similar to those about Shadow IT (OutSystems, 2019) [7]. Low/No-Code development platforms can address the concerns in the same way we discussed above (i.e., the IT department can govern the citizen development applications) (OutSystems, 2019) [7].

If each component or building block in Low/No-Code platforms is secure, reusable, and optimized, the applications built with them should automatically be secure and optimized (OutSystems, 2019) [7]. According to respondents of Outsystems's (2019) survey, 64% of them noticed the scarcity of professional cybersecurity personnel and found that it is hard to hire them [7]. Using Low/No-Code platforms can save time on testing security since IT professionals only need to test the security of building blocks and the platforms themselves (OutSystems, 2019) [7]. Thus, Low/No-Code can not only avoid the security risks but also lessen the heavy testing workload compared to the traditional development process.

####    d.   Maintainability

As Outsystems (2019) estimated, 65% of the application development projects were for maintenance, and the rest were for innovation [7]. Moreover, maintenance activities took approximately 75% of most IT resources of organizations (OutSystems, 2019) [7]. Outsystems (2019) found that organizations that use Low/No-Code development could have more projects for innovation (40%) instead of maintenance compared to those who were not using Low/No-Code development (35%) [7]. Low-Code platforms provide a sole and centralized environment for all aspects of application management, which reduces the complexity and difficulty of application maintenance since critical players in software maintenance (e.g., IT professionals, business professionals) can collaborate on the same platform efficiently (Richardson, C. & Rymer, J. R., 2014) [2]. Moreover, by the nature of Low/No-Code development, maintainers usually only need to maintain a few lines of code, which makes Low/No-Code projects more maintainable (Sanchis et al., 2020) [1].

*B.* LIMITATIONS

Outsystems (2019) presented top reasons that some organizations did not adopt for Low-code development, including Lack of knowledge of Low/No-Code platforms (47%), Worry about vendor lock-in with Low/No-Code providers (37%), Worry about the scalability of Low/No-Code applications (28%), and Worry about the security of Low/No-Code applications (25%) [7]. We expanded several significant concerns about Low/No-code development as we did for advantages.

####    a.   Limited Customizability/Flexibility

The visualized building blocks in Low-Code platforms are pre-implemented and fixed in most cases (Woo, 2020) [6]. Such inflexibility makes the applications less customizable than those developed by traditional coding development (Tay, N., 2021) [16]. It will be difficult and time-consuming to develop complicated or customized features or functionalities that are not provided on the Low-Code platforms (Tay, N., 2021) [16]. Implementing these desired features using codes and integrating them into Low/No-Code applications is an approach, but it lacks consistency and efficiency (Brocoders Company, 2021) [17]. Low/No-Code platforms usually outperform the traditional development process in implementing simple applications where the predefined components address common needs or processes well (Sarabyn, K., 2021) [18]. However, when it comes to projects such as highly customized applications, data science models, or data science workflows, Low/No-Code platforms are not customizable enough for these tasks (Sarabyn, K., 2021) [18].

####    b.   Limited Scalability

Most of the current Low/No-Code platforms are mainly used to develop small-scale applications, while they are seldom used for large-scale, complex, or crucial business applications due to their limited scalability (Sanchis et al., 2020) [1]. According to Rymer and Richardson (2015), the average runtime scale of applications reported by Low-Code platform providers was between 200 and 2,000 concurrent users [14].

####    c.   Security Concerns

Since most Low/No-Code platform users hardly do or cannot customize the applications, they must completely trust that the services do not generate vulnerabilities that cause bugs or data leaks (Oltrogge et al., 2018) [12]. For example, Mobincube, a paid Low-Code service, tracked users silently through Bluetooth low energy beacon without clearly declaring this in the terms and conditions (Oltrogge et al., 2018) [12]. If organizations are dependent on their Low/No-Code platform vendors, their data might be vulnerable to data breaches since data security and source code are not fully controlled by organizations (Tay, N., 2021; Oltrogge et al., 2018) [12] [16]. Moreover, if the platform vendors wind up, there will not be further security updates, and organizations cannot fix new security flaws later (Tay, N., 2021) [16].

####    d.   Vendor lock-in

Warren (2018) defined vendor lock-in as a customer's dependency on a vendor for their products and services, which makes it difficult for the customer to switch to another vendor [19]. In our context, the concern is that organizations will have vendor lock-in with Low/No-Code platform providers. As an organization invests more in an individual Low/No-Code platform provider, the organization will cost more and be more complicated for switching to another platform (Tay, N., 2021) [16].

### III.   CURRENT LOW/NO-CODE DEVELOPMENT PLATFORMS

In the Low/No-Code development industry, besides many emerging LNCD start-ups, tech companies like Microsoft, Alibaba, Salesforce, and Oracle have their own Low/No-Code development platforms (Woo, 2020) [6]. In this section, we analyze several major LNCD platforms in the market for their usage, features, and impacts on digital transformation. According to the study of Kulkarni (2021) on industry demographics of top LNCD platform vendors, we noticed that 20% of Microsoft users were in the manufacturing industry, 40% Appian users in the financial services industry, and 32% Mendix & 30% Salesforce users in services industry [9]. It shows that Manufacturing, Finance, and Services are the three primary industries that utilize LNCD platforms (Kulkarni, 2021)

[9]. For the geographic demographics data of those vendors, 49% Appian and 52% users were in North America, 63% Mendix and 50% Outsystems users were in Europe, Middle East, & Africa, and 37% & 26% Microsoft users were in North America & Asia/Pacific, respectively (Kulkarni, 2021) [9].

Outsystems established the LNCD platform market (Outsystems, 2019) with over 1200 organizations in 52 countries using their platform (Alsaadi et al., 2021) [7][11]. Outsystems focuses on helping customers develop business applications that automate business processes for nimble and continual development & delivery (Alsaadi et al., 2021) [11]. Outsystems platform leverage AI and ML to generate suggestions, automation, and validation of applications developed on the platform (Alsaadi et al., 2021) [11].

Microsoft PowerApps is the LNCD platform provided by tech giant Microsoft. It is integrated with many Microsoft services such as Microsoft Office 365, Microsoft Azure, Microsoft Teams, and so on. It can be used for developing automated business applications by connecting Microsoft services in a workflow (Alsaadi et al., 2021) [11].

Salesforce provides a cloud based LNCD platform that uses Lightning Framework to build business applications more efficiently at a lower cost (Alsaadi et al., 2021) [11]. It gives many customization options for users, such as themes and colours, and those customizations can be reused, making company branding more straightforward without coding (Alsaadi et al., 2021) [11]. It was named a leader for LNCD platforms by Gartner in 2019 (Duffy, 2019) [5].

The LNCD platform of Mendix accelerates application development as other LNCD platforms do. It is outstanding among competitors because it values business & IT cooperation for refining business logic (Alsaadi et al., 2021) [11]. The platform divides the design environment for citizen developers (an environment called Mendix Studio) and professional developers (Mendix Pro) for better collaborations between departments by providing suitable tools for people in different departments. Stakeholders can also cooperate with developers on the platform to provide in-time feedback to boost development and innovation (Alsaadi et al., 2021) [11]. It also has a built-in AI assistant that checks errors and gives suggestions regarding the application's quality, maintainability, and scalability (Alsaadi et al., 2021) [11]. Such AI and ML integration in the platform can give novice developers real-time feedback to better learn the platform and application development.

Appian is smaller than many contestants in scale, but some of its customers are government agencies (Alsaadi et al., 2021) [11]. It is considered for the capabilities of building complicated business processes and applications that require advanced automation and analysis (Alsaadi et al., 2021) [11]. It also has an AI & ML guide integrated to give users immediate suggestions during the development.

We notice that most platforms integrated AI and ML technologies to give real-time recommendations during the development for making up some users' lack of programming skills. The LNCD platforms providers with AI are usually more competitive than others (Alsaadi et al., 2021) [11] since it enhances user experience and reduces concerns about LNCD platform usability. Though current LNCD platforms and technology have defects and limitations (Woo, 2020) [6], we can see the industry is rising, and they are conducive for application development in many cases.

IV. PROSPECT IMPACT ON SOFTWARE DEVELOPMENT

Chris Wanstrath, the CEO of GitHub, once said, "The future of coding is no coding at all." (Peterson, 2017) [21]. Chris Wanstrath claimed that automating the whole development process can make coding easier so that people can focus on the high-level strategies, designs, and architectures of software (Peterson, 2017) [21]. GitHub has plentiful code repositories and troubleshooting forums (Peterson, 2017) [21], which could be used for training AI & ML that gives real-time coding suggestions and error checking. Moskal (2021) argued that Low/No-Code programming is deemed as the fourth generation of programming, and it is also known as declarative programming [13].

However, Thomas Stiehm, the CTO of Coveros, had a contradictory opinion "Low-code is not the future of code. It certainly has a place in the future and will be leveraged to make many applications. It will not replace other ways of creating software because low code breaks down when the solution's complexity increases. We saw the same thing with Visual Basic in the '90s. VB was valuable, and a lot of software was written in VB. In the end, it was the complexity required by some applications that caused VB to break down and no longer be a good solution. Low code will be the same." (Brocoders Company, 2021) [17].

It is true that Low/No-Code development has its limitation and defects, but the analogy between Low/No-Code and VB is inappropriate since VB is an actual product, but Low/No-Code is a high-level concept. People can build their own variation upon a concept but not upon an existing product. For the argument that Low/No-Code cannot create high complexity solutions, we have seen companies such as Appian, Mendix, and Outsystems manage the complex applications well and innovatively. We have also seen the utilization of start-of-the-art AI & ML technologies make LNCD platforms more intelligent and more robust. The defects we discussed before, such as limited scalability and security concern, they can be addressed as LNCD platforms become robust: (1) limited scalability (Warren, 2018) [8]: both runtime scalability (the capability to increase the capacity of deployed LNC apps and to provide fast user experience for both large numbers of users and heavy computation) and dev-time scalability (the capability to use LNC for numerous business requirements, projects, and developers) are proven to be solved in Outsystems' platforms; (2) security concern (Warren, 2018) [20]: Warren argued that if LNCD platforms are robust & well-structured, Low/No-Code applications should be secure by its nature (just as we discussed in Benefits sub-section). Therefore, for Low/No-Code to be best utilized for digital transformation or even to change future software development processes, LNCD platforms vendors must improve the flexibility, scalability, security, and other limitations so that people can accept this new technology without concern.

## V. Conclusion

Organizations' needs for digital transformation provide a stage for Low/No-Code development to show its worth. As digital transformation proceeds, the business application requirements will become more complex and specific. Though LNC has benefits that allow organizations to respond to trends in the industry agilely, they need more flexibility and customizability on the LNCD platforms. LNC provides secure building blocks that can be used for creating secure applications, but data breaches and lack of access to source code raise security concerns. Although there is little or no code to maintain and maintain LNC applications should be easy, vendor lock-in is a problem when the providers no longer support the platforms. Such contradictions show that LNCD could be conducive and show space for LNC technology to improve.

After analyzing current leading LNCD platforms in the market, we found that AI & ML is a key to improving current LNC technology. It is also a key for Low-Code converting to No-Code. As Chris Wanstrath prospected, in the future, people can focus on high-level software prototyping and designing instead of writing lines of code and wasting much time on implementation details. Such a future will come when Low/No-Code technology has wonderfully collaborated with robust coding AI & ML technologies. Thomas Stiehm compared LNC with VB and wanted to conclude the future of LNC should be the same as the one of VB. The conclusion is superficial and unlikely to happen (at least not in the way Thomas predicted). LNC platforms have many potential users, including children, students, professional developers, and non-IT-professional employees. Since it has such a wide range of users, the community should also be large-scale. User community ensures that it will be constantly refined and improved for users' needs. VB did not have such a wide range of users and use-cases, so LNC is not in the old path of VB.

Low/No-Code Development emerged only a few years ago, the adoption rate is rapidly growing, which indicates an optimistic prospect. Researchers around the world should do more research and assessments to solve the limitations and issues with current LNC technology. Organizations can cooperate with LNC platform providers to improve their platforms. It is not only for the providers but also for organizations to have a better user experience. LNC platform vendors should continuously research and address limitations by studying competitors' work and seeking to combine the latest technologies with the platforms.

In conclusion, based on our analysis in each section, Low/No-Code Development will play a crucial role in digital transformation and cause a turnaround in the software development industry.